# Tuning efficiency and sensitivity of guided resonances in photonic crystals and quasi-crystals: a comparative study

**Marco Pisco,[1] Armando Ricciardi,[2] Ilaria Gallina,[3] Giuseppe Castaldi,[3] Stefania Campopiano,[2] Antonello Cutolo[1], Andrea Cusano[1,\*], and Vincenzo Galdi[3]**

[1] *Optoelectronic Division, Dept. of Engineering, University of Sannio, Corso Garibaldi 107, I-82100 Benevento, Italy*
[2] *Dept. for Technologies, University of Naples "Parthenope," Centro Direzionale, Isola C4, I-80143, Naples, Italy*
[3] *Waves Group, Dept. of Engineering, University of Sannio, Corso Garibaldi 107, I-82100 Benevento, Italy*
*a.cusano@unisannio.it*

**Abstract:** In this paper, we present a comparative study of the tuning efficiency and sensitivity of guided resonances (GRs) in photonic crystal (PC) holed slabs based on periodic and aperiodically-ordered unit cells, aimed at assessing the applicability of these important technology platforms to ultra-compact optical sensors and active devices. In particular, with specific reference to square-lattice periodic PCs and aperiodically-ordered Ammann-Beenker photonic quasi-crystals, we study the effects of the hole radius, slab thickness, and refractive index on the GR sensitivity and tunability with respect to variation in the hole refractive index. Finally, we carry out a theoretical and numerical analysis in order to correlate the GR shift with the field distribution of the unperturbed (air holes) structures. Our results indicate that the spatial arrangement of the holes may strongly influence the tuning and sensitivity efficiency, and may provide new degrees of freedom and tools for the design and optimization of novel photonic devices for both sensing and telecommunication applications.

## 1. Introduction

Guided resonances (GRs) in photonic crystal (PC) slabs have been widely investigated during the last decade, and significant efforts have been devoted to the study of the underlying physical mechanisms [1] as well as to their experimental observation and characterization [2-5]. Such efforts were mainly motivated by the fascinating potentials that they offer in the light-flow manipulation for sensing and communication applications.

From the phenomenological viewpoint, GRs rely on the coupling among the continuum of free-space *radiative* modes and the discrete set of *leaky* modes supported by a PC slab [1]. This wavelength-selective mechanism provides out-of-plane transmission and reflection spectra exhibiting very sharp spectral variations over a narrow frequency range, also known as Fano resonances [1-7]. Recently, the possibility of exciting GRs in "photonic quasi-crystal" (PQC) slabs based on the Ammann-Beenker (octagonal) aperiodic tiling [8] has been numerically envisaged, opening up new attractive perspectives for high-performance photonic devices and components [9].

Both in PCs and in PQCs, besides the dependence on the direction and polarization of the incident wave [10-11], the structural parameters of the slab also play a key role in determining the GR spectral features, as already demonstrated in some numerical and experimental studies [11-12]. Consequently, it can easily be inferred that the GR-wavelength tuning can be more or less efficiently attained by controlling the constitutive and geometrical parameters of the structure, including the slab refractive index and thickness, the lattice constant, and the filling factor. On the other hand, in view of their power distribution strongly confined in the slab (and yet the capability of coupling with the external radiation), GRs provide an *inherent sensitivity* to the external environment. More specifically, as long as the transmission through the slab is influenced by the refractive index changes in the surrounding environment, the GRs exhibit a wavelength shift [13-14].

Based on these arguments, several ultra-compact GR-based optical devices have been proposed. In 2007, Levi *et al.* [13] proposed the exploitation of GR effects in PCs to develop miniaturized sensors capable of detecting refractive index changes in aqueous solutions. They observed a GR spectral shift in experiments featuring a bulk fluid surrounding the slab, with a wavelength sensitivity of 130nm per refractive index unit (RIU). In 2009, Huang *et al.* [14] integrated the refractive index detection functionality with a nanofluidics system capable of forcing the flow through the slab holes, and achieved a sensitivity enhancement up to a value of 510nm/RIU, basically by improving the interaction between light and fluid in the slab holes. In a numerical study, Song *et al.* [15] proposed to exploit GRs in order to develop optical tunable devices by changing the refractive index of the PC slab, whereas in Shi *et al.* [16] proposed a GR-based biochemical sensor constituted by a monolayer of spherical voids (or silica beads) in a dielectric slab.

Moreover, micromechanically tunable optical filters [17-18] and nanoscale displacement sensors [18-19] have also been proposed and theoretically explored, both basically composed of two PC slabs facing each other. Recently, the same configuration gained further attention in numerical studies aimed at investigating the performance of the "two-faced-slabs structure" for its potential application to high-sensitivity fluorescence-detection-based biosensors [20] and advanced opto-mechanical devices [21]. In spite of the interesting performance attainable for both sensing and communication applications, and because of the difficulty of fabrication, such a complex structure has only recently been implemented in a passive configuration for optical filtering functions [22]. In connection with *active* devices, instead, Kanamori *et al.* [23] fabricated and characterized a micromechanically tunable optical filter based on the evanescent coupling of GR in a single PC slab with a silicon substrate. The resonance wavelength was controlled by changing the gap between the PC slab and the substrate via microelectromechanical actuators.

Against the above background, in this paper, we present a comparative study of GRs in PC and PQC slabs, with the aim of providing a systematic approach to the design and optimization of sensing or active devices and components. In this framework, we carry out a comprehensive parametric study of the tuning efficiency and sensitivity characteristics, as a function of the slab refractive index, thickness, and filling ratio, whose results provide useful indications and tools for a concurrent optimization of the final device. More specifically, for a given slab parameter configuration, we calculate the GR spectral shifts induced by variations of the hole refractive index, for both the PC and PQC cases. Moreover, our study also includes a theoretical and numerical analysis based on the cavity perturbation theory, aimed at predicting the sensitivity and tuning efficiency of the GRs on the basis of the unperturbed (air holes) modal field distributions.

Accordingly, the rest of the paper is laid out as follows. In Sec. 2, we illustrate the problem geometry and the key parameters and observables in our numerical analysis. In Sec. 3, we present the numerical results, and apply the cavity perturbation theory to model the GR wavelength shift in PCs and PQCs. Finally, in Sec. 4, we provide some concluding remarks and hints for future research.

## 2. Problem geometry, parameters, observables, and methodology

In our comparative study, we consider GRs excited in two representative PQC and PC slab configurations.

In connection with PQCs, we consider an aperiodic lattice based on the octagonal Ammann-Beenker tiling made of square and rhombus tiles (see [8] for details). The tiling vertices are assumed as centers of circular holes with refractive index $n_{holes}$ in a dielectric slab with refractive index $n_{slab}$. We indicate with *a* the lattice constant (which, in the considered aperiodic lattice, represents the square/rhombus tile sidelength [8]), with *r* the hole radius, with *t* the slab thickness. As discussed in our previous study on PQCs [8], strictly speaking, we study a *periodic* PC by limiting and replicating the (theoretically endless) aperiodic tiling, with a suitably large supercell [shown in Fig 1(a), with sidelength $L = (4+3\sqrt{2})a$ ].

As a standard *periodic* PC, we consider a square lattice of holes [shown in Fig. 1(b)], and the same notation as above (with the lattice constant *a* now being the period).

In order to meaningfully compare the results concerning the PC vs. PQC case, we choose the geometrical parameters in order to equalize the hole/dielectric volumetric fraction, i.e.,

$$\left(\frac{r}{a}\right)_{PC} = \frac{\sqrt{82}}{4+3\sqrt{2}}\left(\frac{r}{a}\right)_{PQC} \approx 1.1\left(\frac{r}{a}\right)_{PQC}, \qquad (1)$$

keeping the lattice constant fixed, so as to compare different structures with *homologous* geometries.

For both PC and PQC configurations, evidences of GRs in transmission response have been already reported [1, 8, 12], and are not the subject of the present study. Instead, we focus our attention on the dependence of the GR wavelength on the hole refractive index. To this aim, we carry out a preliminary investigation of the band-structures associated with the unit cells in Fig. 1, with particular emphasis on the Γ point of the irreducible Brilluoin zones. This allows systematic calculation of the modal field distributions associated with the selected PC cell and PQC supercell, and determination of their symmetry properties, so as to ascertain their couplability with a normally-incident plane wave [6-9]. In particular, we consider modal solutions exhibiting both *even* (transverse-electric-like) and *odd* (transverse-magnetic-like) symmetry along the *z*-axis, focusing the attention on the lowest-order eigen-modes that exhibit an in-plane symmetry matching that of a normally-incident plane wave.

The modal analysis is carried out by using a commercial software package (RSOFT BandSOLVE® [24]) based on a 3-D plane-wave expansion method. The computational domain includes the slab, lying in the transverse (*x-y*) plane, surrounded by air layers of thickness *7.5a* at each side along the *z* axis, with periodic boundary conditions applied to the domain terminations in all dimensions [24]. The computational domain is discretized via uniform grids along the *x*, *y* and *z* axes. Numerical convergence was achieved for a discretization of at least ten elements per minimum wavelength (per dimension), resulting in a grand total of about $10^6$ elements.

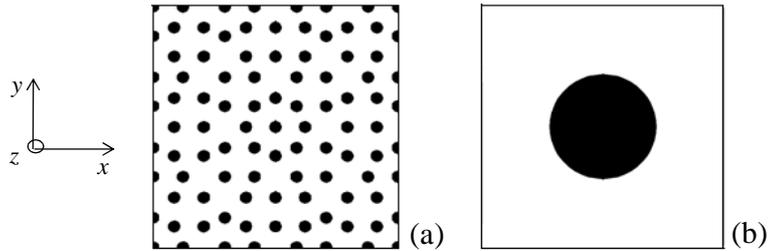

Fig. 1. Lattice geometry in the Cartesian reference system: (a) Octagonal Ammann-Beenker supercell; (b) square lattice cell.

## 3. Numerical results and discussion

*3.1. Parametric analysis*

In our simulations, both for the PQC supercell and the PC cell, we first consider a silicon slab ($n_{slab}$=3.418) surrounded by air. We compute the GR wavelength shift induced by variations of the hole refractive index within the range [1, 1.5], focusing the attention on the GRs corresponding to the lowest-order *z*-even and *z*-odd modes exhibiting an in-plane even symmetry in the electric and magnetic fields, respectively [6-9]. Such study is repeated for different parameter configurations, by varying the hole radius and the slab thickness.

With specific reference to the PQC supercell [cf. Fig. 1(a)], for a given slab thickness $t=0.75a$, we consider three values of the hole radius, $r=0.20a, 0.25a, 0.30a$. Similarly, for a fixed hole radius $r=0.25a$, we consider three values of the slab thickness, $t=0.50a, 0.75a, a$. In Fig. 2, we show the GR wavelength shift for the lowest-order $z$-even (a,b) and $z$-odd mode (c,d), varying the hole radius (a, c) and slab thickness (b, d). The reference wavelength ($\lambda$) values are calculated for each case from the GR normalized frequencies ($a/\lambda$, summarized in table I) by setting the lattice constant; without loss of generality, we set such value to $a=400$ nm.

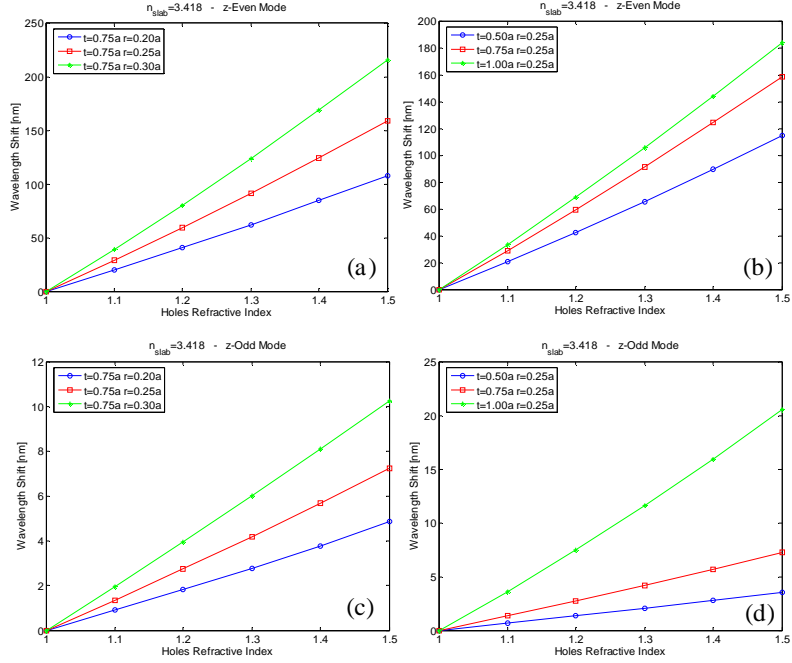

Fig. 2. GR wavelength shift as a function of the hole refractive index in a PQC supercell [cf. Fig. 1(a)] with high-index ($n_{slab}=3.418$) slab for the lowest-order $z$-even (a,b) and $z$-odd (c,d) modes, and different values of the hole radius (a, c) and slab thickness (b, d).

Table I. Normalized GR frequencies in the unperturbed PQC with high index ($n_{slab}=3.418$) slab for the lowest-order $z$-even and $z$-odd modes, for various values of the hole radius and slab thickness.

| Normalized Frequencies $a/\lambda$ | HIGH INDEX ($n_{slab}=3.418$) | | | | | | |
|---|---|---|---|---|---|---|---|
| | $z$-even | | | | $z$-odd | | |
| | $t \backslash r$ | 0.20a | 0.25a | 0.30a | $t \backslash r$ | 0.20a | 0.25a | 0.30a |
| PQC | 0.50a | | 0.091 | | 0.50a | | 0.117 | |
| | 0.75a | 0.076 | 0.081 | 0.087 | 0.75a | 0.115 | 0.115 | 0.116 |
| | 1.00a | | 0.074 | | 1.00a | | 0.109 | |

As evident from Fig. 2, in all cases, the GR wavelength increases with the hole refractive index. The results reveal also that broader hole radii, as well as thicker slabs, yield larger wavelength shifts for the same variations of the hole refractive index. This is attributable to the increased volume of interaction between the GR field distribution and the hole refractive index, attainable by increasing *r* or *t*. This aspect will be better elucidated in Sec. 3.3 below.

For the PC cell [cf. Fig. 1(b)], the geometrical parameters are chosen according to Eq. (1), equalizing the hole/dielectric fraction with respect to the PQC case above. This results, for a slab thickness *t=0.75a*, in hole radii *r=0.22a, 0.275a, 0.33a* (maintaining the same lattice constant *a*). Similarly, for a hole radius *r=0.275a*, we consider the same slab thickness values (*t=0.50a, 0.75a, a*) as above. The corresponding results are summarized in Fig. 3 and table II. By comparison with the PQC case above (cf. Fig. 2 and table I), similar qualitative considerations hold in connection with the trends of the GR wavelength shift. However, a *non-linear* behavior of the wavelength shift (as a function of the hole refractive index) is now evident.

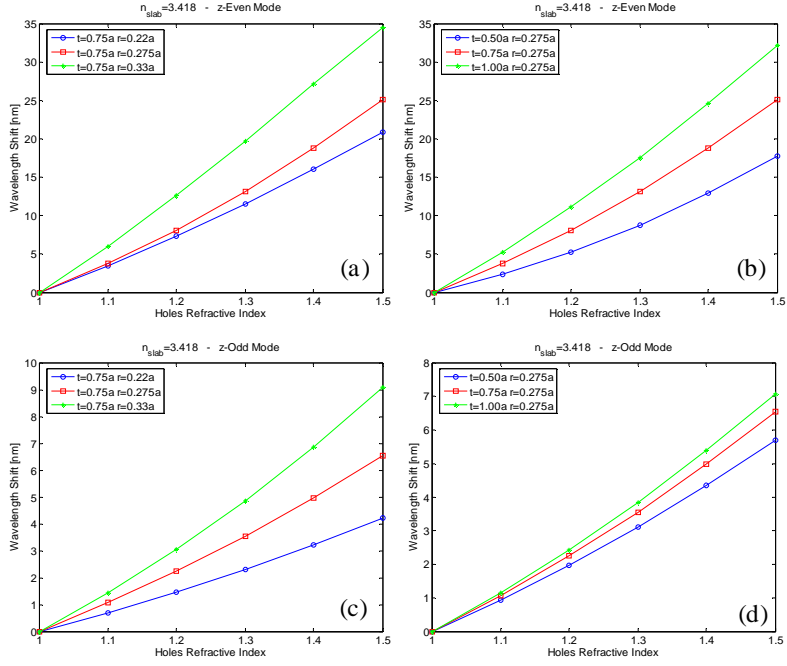

Fig. 3. As in Fig. 2, but for the periodic PC case [cf. Fig. 1(b)].

**Table II. As in table I, but for the periodic PC case [cf. Fig. 1(b)].**

| Normalized Frequencies $a/\lambda$ | HIGH INDEX ($n_{slab}$=3.418) | | | | | | | |
|---|---|---|---|---|---|---|---|---|
| | *z*-even | | | | *z*-odd | | | |
| | *t \ r* | *0.22a* | *0.275a* | *0.33a* | *t \ r* | *0.22a* | *0.275a* | *0.33a* |
| PC | *0.50a* | | **0.421** | | *0.50a* | | **0.435** | |
| | *0.75a* | **0.379** | **0.400** | **0.440** | *0.75a* | **0.365** | **0.377** | **0.393** |
| | *1.00a* | | **0.390** | | *1.00a* | | **0.353** | |

In order to investigate the effect of the slab refractive index on the tuning efficiency, the same studies have been carried out by considering a slab refractive index $n_{slab}=1.58$. Such a low index was considered in view of the increasing interest in photonic polymeric materials for sensing and active devices and, more specifically, in light of the recent developments in the fabrication of periodic patterns using soft materials [25].

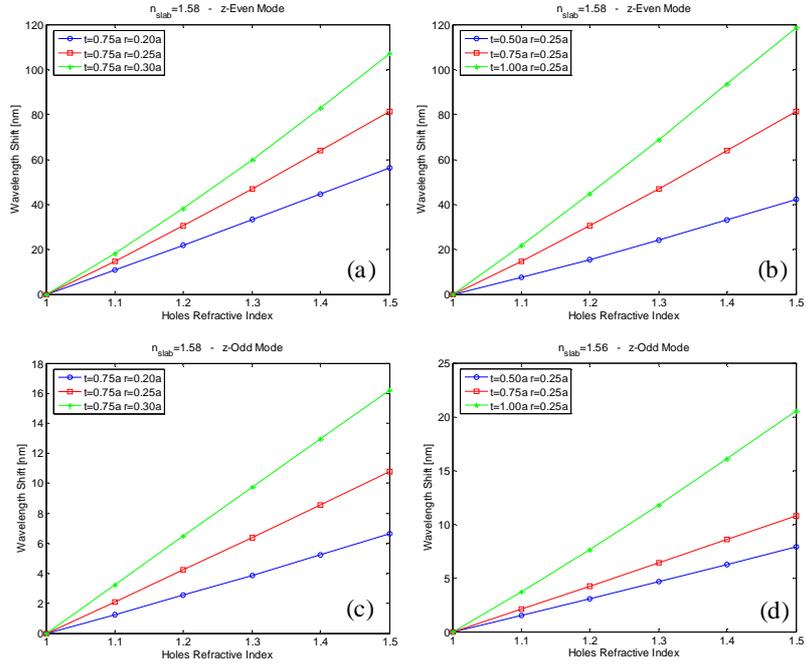

Fig. 4. As in Fig. 2, but for a low-index ($n_{\text{slab}}$=1.58) slab.

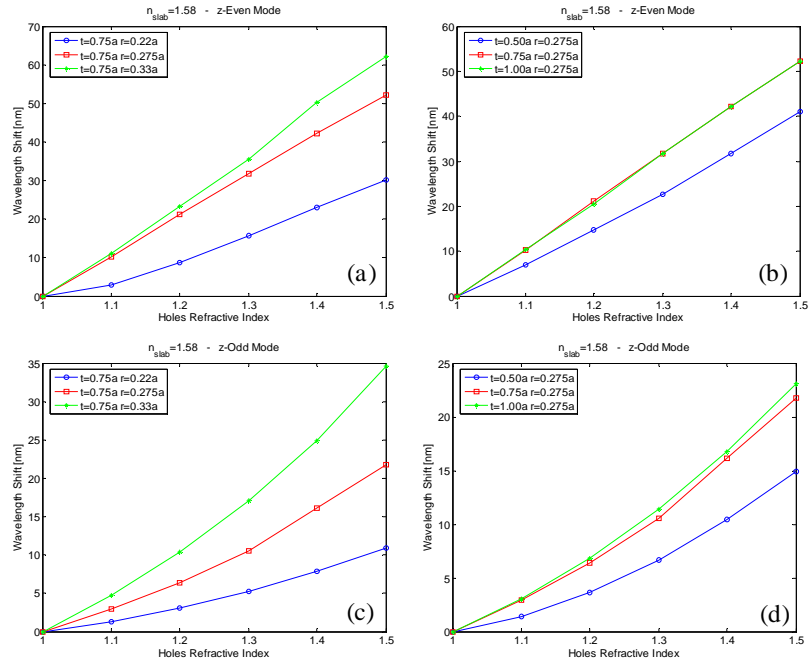

Fig. 5. As in Fig. 3, but for a low-index ($n_{\text{slab}}$=1.58) slab.

The corresponding numerical results, in terms of GR wavelength shift, are shown in Figs. 4 and 5, with reference to the PQC and PC case, respectively, while table III summarizes the normalized reference frequencies for both cases. In this low-index slab case, a non-linear trend in the GR wavelength shift can be observed for both the PC and PQC configurations.

It is worth noting that, in general, the wavelength shifts tend to be larger for higher values of *r* and *t*. However, as evident from Fig. 5 (b, d), in the PC case the increase of the wavelength shift saturates for increasing values of *t*. This effect is attributable to a strong localization of the field distribution (electric field for *z*-even, and magnetic field for *z*-odd) within the PC slab, such that a further increase of the thickness leaves the field distribution essentially *unperturbed*.

Table III. As in tables I and II, but for a low-index ($n_{slab}$=1.58) slab.

| Normalized Frequencies $a/\lambda$ | LOW INDEX ($n_{slab}$=1.58) | | | | | | | |
|---|---|---|---|---|---|---|---|---|
| | *z*-even | | | | *z*-odd | | | |
| PQC | *t* \ *r* | 0.20a | 0.25a | 0.30a | *t* \ *r* | 0.20a | 0.25a | 0.30a |
| | 0.50a | | 0.118 | | 0.50a | | 0.120 | |
| | 0.75a | 0.115 | 0.117 | 0.118 | 0.75a | 0.119 | 0.119 | 0.119 |
| | 1.00a | | 0.114 | | 1.00a | | 0.118 | |
| PC | *t* \ *r* | 0.22a | 0.275a | 0.33a | *t* \ *r* | 0.22a | 0.275a | 0.33a |
| | 0.50a | | 0.806 | | 0.50a | | 0.808 | |
| | 0.75a | 0.745 | 0.756 | 0.797 | 0.75a | 0.738 | 0.756 | 0.780 |
| | 1.00a | | 0.756 | | 1.00a | | 0.726 | |

*3.2. GR sensitivity and tuning efficiency in PCs and PQCs*

In order to address a *quantitative* comparison of the previously analyzed structures, we define the following "sensitivity–tuning efficiency" observable:

$$S(n_{holes}) = \frac{\partial \lambda}{\partial n_{holes}}. \qquad (2)$$

As previously mentioned, the GR wavelength shifts have been evaluated with respect to variations of the refractive index of the holes (arranged in periodic or aperiodic fashion), and maintaining the refractive index of the surrounding environment unchanged. This configuration is suitable for biological and chemical sensing applications, where the GR sensitivity with respect to variations of the hole refractive index represents the transduction principle of a PC- or PQC-based sensor featuring a sensitive material integrated within the slab holes. The presence of a specific sensitive material in the slab holes, undergoing a refractive index change via specific target molecules adsorption [26], improves the overall sensing performance since it allows detection of low target analyte concentrations (which would be unable to change the bulk refractive index). From a dual perspective, in *active* devices, the same principle can be effectively exploited for "functionalization" purposes, by actuating a material infiltrated in the slab holes.

In Figs. 6 and 7, we show the sensitivity-tuning efficiency corresponding to the high-index slab cases in Figs. 2 (PQC) and 3 (PC), respectively. It can be observed that the efficiency is strictly monotonically increasing. The maximum efficiencies observed, which therefore occur for $n_{holes}$=1.5, are reported in table IV.

As it could already be deduced from Figs. 2 and 3, the above results confirm that, in both the PQC and PC cases, the sensitivity-tuning efficiency increases with *r* and *t*. Moreover, it can be noted from table IV that, in the high-index case, the sensitivity of the GRs associated to *z*-even modes tend to be higher than those associated to *z*-odd modes. More specifically, for *z*-even modes, the GR sensitivities observed in the PQC case tend to be higher than their PC counterparts, while the sensitivities related to *z*-odd modes are basically comparable.

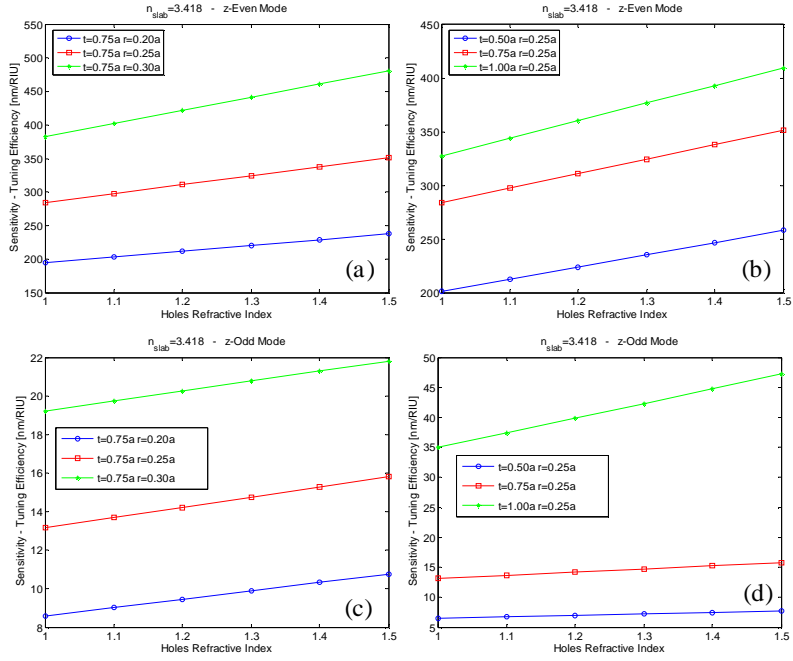

Fig. 6. Sensitivity–tuning efficiency as a function of the hole refractive index in a PQC supercell [cf. Fig. 1(a)] with high index ($n_{slab}$=3.418) slab for the lowest-order $z$-even (a,b) and $z$-odd (c,d) modes, and different values of the hole radius (a, c) and slab thickness (b, d).

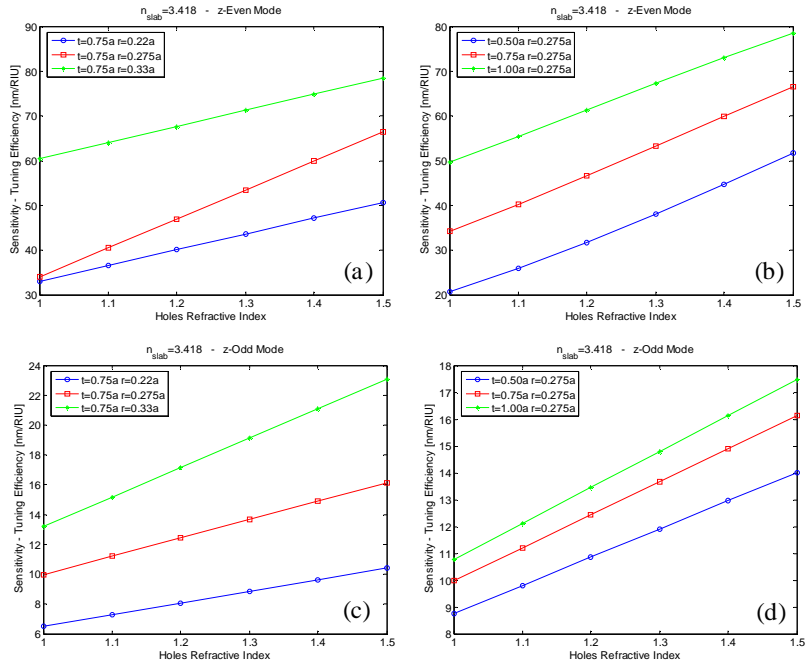

Fig. 7. As in Fig. 6, but for the periodic PC case [cf. Fig. 1(b)].

**Table IV. Maximum sensitivity-tuning efficiencies observed (for $n_{\text{holes}}$=1.5) in the PQC and PC cases (cf. Figs. 6-9).**

| Maximum Sensitivity--Tuning Efficiency (at $n_{\text{holes}}$=1.5) [nm/RIU] | HIGH INDEX ($n_{\text{slab}}$=3.418) | | | | | | | | LOW INDEX ($n_{\text{slab}}$=1.58) | | | | | | | |
|---|---|---|---|---|---|---|---|---|---|---|---|---|---|---|---|---|
| | z-even | | | | z-odd | | | | z-even | | | | z-odd | | | |
| **PQC** | t \ r | 0.20a | 0.25a | 0.30a | t \ r | 0.20a | 0.25a | 0.30a | t \ r | 0.20a | 0.25a | 0.30a | t \ r | 0.20 | 0.25a | .25a |
| | 0.50a | | 258.1 | | 0.50a | | 7.7 | | 0.50a | | 96.2 | | 0.50a | | 16.9 | |
| | 0.75a | 237.7 | 351.4 | 480.7 | 0.75a | 10.8 | 15.8 | 21.8 | 0.75a | 117.0 | 179.5 | 253.2 | 0.75a | 14.2 | 22.3 | 32.5 |
| | 1.00a | | 409.4 | | 1.00a | | 47.2 | | 1.00a | | 258.2 | | 1.00a | | 46.0 | |
| **PC** | t \ r | 0.22a | 0.275a | 0.33a | t \ r | 0.22a | 0.275a | 0.33a | t \ r | 0.22a | 0.275a | 0.33a | t \ r | 0.22a | 0.275a | 0.33a |
| | 0.50a | | 51.0 | | 0.50a | | 14.0 | | 0.50a | | 97.1 | | 0.50a | | 49.0 | |
| | 0.75a | 50.6 | 66.5 | 78.4 | 0.75a | 10.4 | 16.1 | 23.1 | 0.75a | 86.4 | 105.1 | 138.6 | 0.75a | 32.8 | 63.3 | 99.0 |
| | 1.00a | | 79.0 | | 1.00a | | 17.5 | | 1.00a | | 105.1 | | 1.00a | | 66.4 | |

Results pertaining to the low-index case are shown in Figs. 8 and 9, with the maximum sensitivities still reported in table IV.

In this case, it can be observed from the table IV that the efficiencies in the PQC case tend to be higher than their PC counterpart for z-even modes, whereas they tend to be lower for z-odd modes. Furthermore, the efficiencies observed for z-even modes tend to be higher than the those observed for z-odd modes, in both the PQC and PC cases.

By comparing the overall results obtained with high- and low-index slabs, for PQCs and PCs, counterintuitive differences in the behavior of the GR wavelength-shift as a function of the hole refractive index can be noted. As matter of fact, the GRs associated to z-even modes in PQCs tend to be more sensitive in the high-index slab case than in low-index slab case, while the contrary (i.e., GRs associated to even modes more sensitive in the low-index slab case) happens in the PC case.

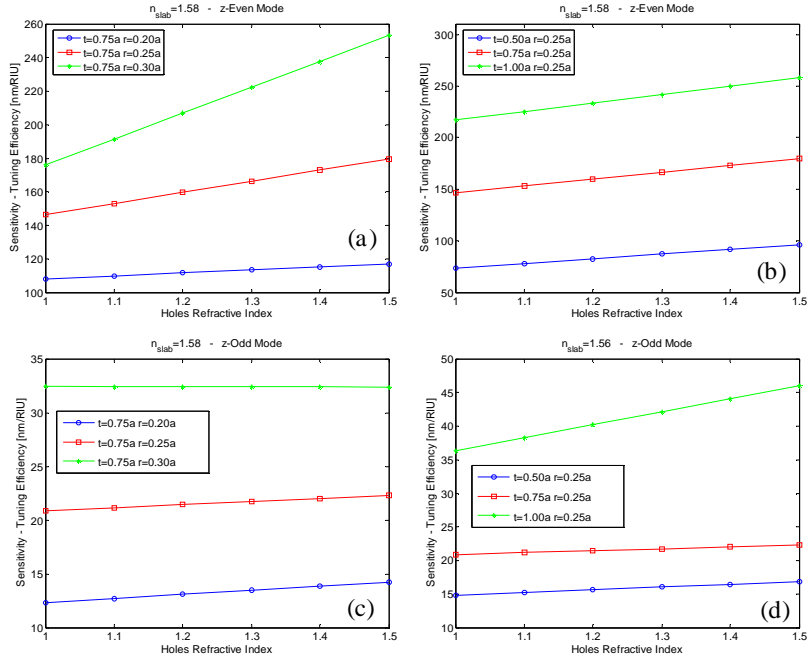

Fig. 8. As in Fig. 6, but for a low-index ($n_{\text{slab}}$=1.58) slab.

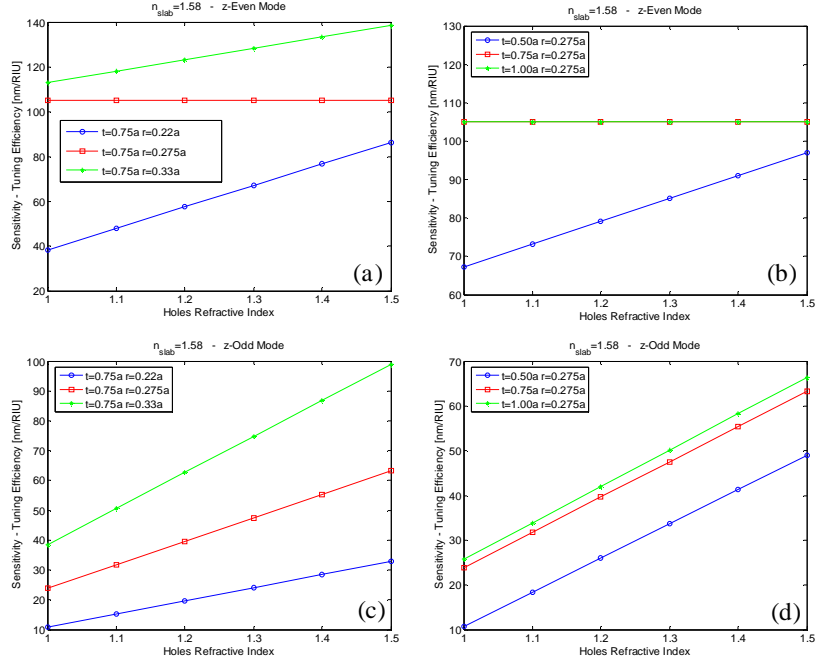

Fig. 9. As in Fig. 7, but for a low-index ($n_{\text{slab}}$=1.58) slab.

*3.3. Perturbative model for GRs in PCs and PQCs*

In view of the above illustrated complex dependence of the GR sensitivity on the slab refractive index, mode type (*z*-even or *z*-odd), and lattice geometry (PC or PQC), in what follows, we try to gain further insight in the relationship between the wavelength shift and the modal field distribution of the unperturbed ($n_{\text{holes}}$=1) structures. In particular, we apply the perturbation theory, originally developed in connection with microwave cavity tuning [27], to model the dependence of GRs wavelength shift on the hole refractive index. In this framework, generalizing the results in [27, Sec. 7.3] to the case of a *fully dielectric* cavity composed of the unperturbed ($n_{\text{holes}}$=1) PC or PQC slab surrounded by the air layers along the *z*-axis and terminated by periodic boundary conditions in all directions, the GR wavelength shift due to a variation in the hole refractive index (i.e., $n_{\text{holes}}\neq 1$) can be *rigorously* related to the unperturbed $(E_0, H_0)$ and perturbed $(E, H)$ electric and magnetic modal field distributions as follows:

$$\Delta\lambda = \lambda_0 \frac{\left(n_{\text{holes}}^2 - 1\right) \iiint\limits_{\text{holes}} E(r) \cdot E_0^*(r) \, dr}{\iiint\limits_{\text{cavity}} \left[n^2(r) E(r) \cdot E_0^*(r) + \eta_0^2 H(r) \cdot H_0^*(r)\right] dr}, \quad (3)$$

where $\lambda_0$ and $\eta_0$ indicate the unperturbed GR wavelength and the vacuum characteristic impedance, respectively, $^*$ denotes complex conjugation, and

$$n(r) = \begin{cases} n_{\text{slab}}, & \text{in the slab,} \\ 1, & \text{in the holes and background.} \end{cases} \quad (4)$$

Note that, albeit *exact*, Eq. (3) requires the complete knowledge of the *perturbed* modal field distribution, and so its practical applicability is very limited. However, in the limit of *small*

*perturbations* ($n_{holes}\approx 1$), where one can assume $\mathbf{E} \approx \mathbf{E}_0$, $\mathbf{H} \approx \mathbf{H}_0$, it can be simplified to a form that involves only the electric modal field distribution of the *unperturbed* structure, viz.,

$$\Delta\lambda \approx \lambda_0 \frac{\left(n_{holes}^2-1\right)\iiint_{holes}\left|\mathbf{E}_0(\mathbf{r})\right|^2 d\mathbf{r}}{2\iiint_{cavity} n^2(\mathbf{r})\left|\mathbf{E}_0(\mathbf{r})\right|^2 d\mathbf{r}}, \tag{5}$$

where use has been made of the equality between the resonant unperturbed magnetic and electric energy densities.

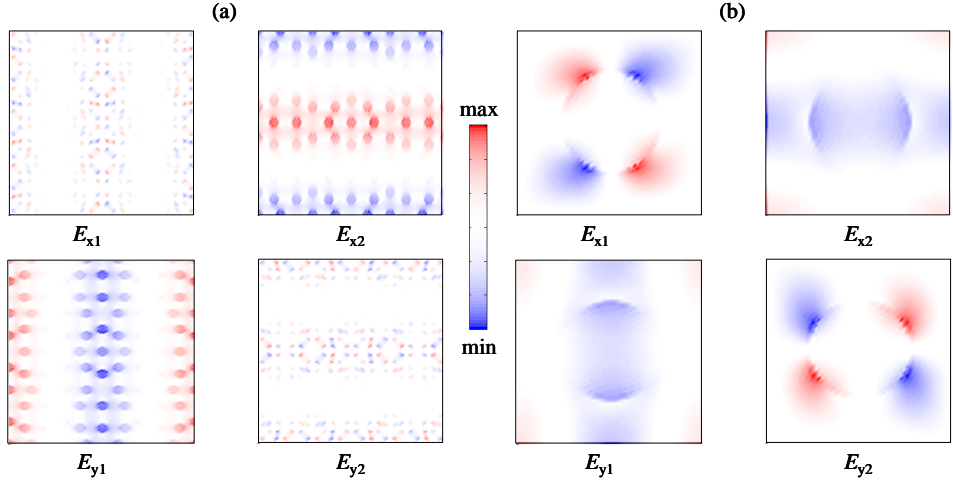

Fig. 10. Electric field amplitudes (at the slab center *x-y* plane) for *x-* and *y-*components of the lowest order *z*-even degenerate modes, in the PQC (a) and PC (b) case, with high-index slab.

From the physical viewpoint, Eq. (5) states that the GR wavelength response to variations of the holes refractive index strongly depends on the electric field concentration in the slab holes. In order to assess to what extent such model is able to capture and explain the observed differences among GRs associated to *z*-even and *z*-odd modes, in PCs and PQCs with high- and low-index slabs, we compare the wavelength shift calculated according to Eq. (5) with the corresponding results from our previous *full-wave* study. For brevity, we restrict this comparison to the configuration that was observed to provide the highest sensitivity, i.e., the lowest-order *z*-even mode in the PQC supercell with high-index slab ($n_{slab}=3.418$). In particular, we refer to the configuration with hole radius $r=0.25a$ and slab thickness $t=0.75a$, whose modal field distributions have already been extensively investigated in [8]. In Fig. 10(a) we recall the in-plane electric-field *x-* and *y-* components of the two degenerate modes ($E_{xj}$ and $E_{yj}$, with $j=1,2$), exhibiting the mirror symmetries that allow the coupling with a normally-incident (*y-* or *x-*polarized, respectively) plane wave. Similarly, in Fig. 10(b), we show the PC counterparts, pertaining to a structure with hole radius $r=0.275a$ [cf. Eq. (1)].

In Fig. 11, the GR wavelength shifts resulting from our previous full-wave modal analysis are compared with those predicted from the numerical implementation of the perturbative model in Eq. (5). A substantial agreement in the trends can be observed, thereby confirming that the previously highlighted differences between PCs and PQCs in terms of sensitivity-tuning efficiency are actually captured by the simple perturbative model in Eq. (5).

To sum up, by applying simple perturbation arguments to GRs in PC and PQC slabs, we ascertained that the GR sensitivity to variations in the hole refractive index is basically dictated by the electric field concentration in the holes. Therefore, the counterintuitive differences emerged in our parametric studies are basically attributable to different field distributions in the slab holes.

The above analysis, besides offering a physical insight in the observed differences, also suggests useful prediction/design strategies which rely on the field distribution of the unperturbed structure. In this framework, the numerical results suggest also that **t**he spatial arrangement of the holes (and, in particular, the *aperiodic* arrangement in PQC-type supercells) represents an important degree of freedom to increase the interaction with the material infiltrated in the holes.

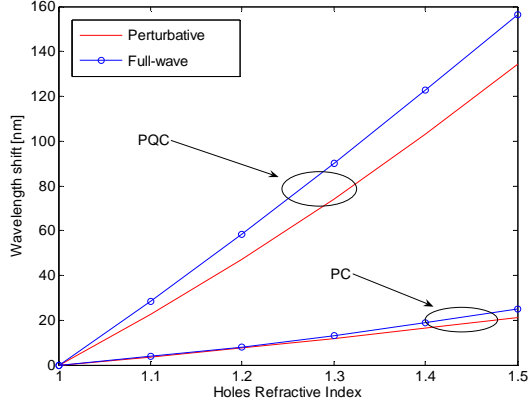

Fig. 11. GR wavelength shift calculated via full-wave modal analysis and perturbative model [cf. Eq. (5)], pertaining to the lowest order *z*-even mode, in the PC ($r$ =0.275$a$) and PQC ($r$ =0.25$a$) case, with a high-index ($n_{slab}$=3.418) slab of thickness $t$=0.75$a$.

## 4. Conclusions

In this paper, we have presented a comparative numerical study of the GR properties in PC and PQC slabs constituted by arrangements of holes (with periodic and aperiodically-ordered supercells, respectively) in a dielectric host medium. The analysis has been focused on the GR dependence on the hole refractive index, and comparisons have been drawn in terms of GR wavelength sensitivity and tuning efficiency. Such sensing/tuning scheme is representative of a device in which a sensitive/active material is infiltrated in the slab holes.

Our numerical results highlighted the dependence of the GR sensitivity-tuning efficiency on the physical and geometrical parameters characterizing the patterned slab, as well as on the nature of the GR modes (*z*-even or *z*-odd). Specifically, for the parameter settings considered, the best performance was achieved with GRs associated to the lowest-order *z*-even mode in a PQC slab with $n_{slab}$=3.418. It is important to remark that, although in the specific context investigated in this work, PQC-based GRs turn out to exhibit higher sensitivity than their periodic counterpart, a general rule cannot be derived at this stage, since a deeper analysis is necessary, involving different tiling geometries (e.g., Penrose, dodecagonal, etc.) for the PQC case and different symmetries (e.g., hexagonal) for the periodic case.

Finally, by applying simple perturbation arguments, we have investigated the dependence of the GR wavelength shift on the modal field distribution of the unperturbed (air holes) structure. Overall, our results indicate that the GR sensitivity to variations in the hole refractive index is basically dictated by the electric field concentration in the holes, and that, within the investigated context, the hole arrangement in the PQC case provides a significant improvement of the light-matter interaction via a stronger field localization in the holes. Moreover, the agreement between the numerical results and those obtained through the perturbative approach suggests a systematic approach for the design and optimization of highly sensitive/active devices and components by acting on the hole spatial arrangement in the unperturbed structures.

Current and future studies are aimed at the exploration of other PC and PQC geometries (based on aperiodic tilings or substitutional sequences), as well as at the experimental verification of the results and their application to the design of high-performance sensing/active devices.

**Acknowledgments**

The kind assistance of Prof. M. N. Armenise (Polytechnic of Bari, Italy) in the modal analysis is gratefully acknowledged.